# Acceleration of trapped particles and beams

Er'el Granot[(1)] and Boris Malomed[(2)]

[(1)]*Department of Electrical and Electronics Engineering, Ariel University Center of Samaria, Ariel, Israel*

[(2)]*Department of Physical Electronics, School of Electrical Engineering, Faculty of Engineering, Tel Aviv University, Tel Aviv 69978, Israel*

**Abstract**

The dynamics of a quantum particle bound by an accelerating delta-functional potential is investigated. Three cases are considered, using the reference frame moving along with the δ-function, in which the acceleration is converted into the additional linear potential. (i) A stationary regime, which corresponds to a resonance state, with a minimum degree of delocalization, supported by the accelerating potential trap. (ii) A *pulling scenario*: an initially bound particle follows the accelerating delta-functional trap, within a finite time. (iii) The *pushing scenario*: the particle, which was initially localized to the right of the *repulsive* delta-function, is shoved to the right by the accelerating potential. For the two latter scenarios, the life time of the trapped particle, and the largest velocity to which it can be accelerated while staying trapped, are found. Analytical approximations are developed for the cases of small and large accelerations in the pulling regime, and also for a small acceleration in the stationary situation, and in the regime of pushing. The same regimes may be realized by Airy-like planar optical beams guided by a narrow bending potential channel or crest. Physical estimates are given for an atom steered by a stylus of a scanning tunneling microscope (STM), and for the optical beam guided by a bending stripe.

## 1. Introduction.

Recently, vast improvements to techniques of manipulating micro- and nano-particles have been reported. Single particles can be picked up and translated by the stylus of the scanning tunnel microscope (STM) [1-6] and by laser beams [7-9]. These advances suggest to consider limitations of the techniques. One of the issues is acceleration of a quantum particle to a given velocity, as the accelerated particle may be excited energetically beyond the level of the holding potential. Therefore, unless the acceleration is very slow, there is a finite probability of loosing the particle. In applications (the transfer of individual atoms), the acceleration may be followed by deceleration of the driven atom. Due to the time-reversal symmetry, the deceleration problem is actually equivalent to the case of the acceleration.

Similar scenarios can be implemented in a completely different physical setting, *viz.*, the guidance of optical beams in planar media by a narrow bent stripe (trapping channel or repulsive barrier), which is a counterpart of the delta-functional potential in optics. In this case, it is natural to expect that the "accelerating" beams will resemble the Airy modes, which have recently drawn considerable attention in optics [10]. The formulation of the problem given below is interpreted in terms of both quantum mechanics and optics.

The objective of this work is to investigate various aspects of the acceleration problem (in the framework of the linear theory). The model is formulated in a detailed form in Section 2, using the reference frame moving along with the accelerating delta-function. Then, we consider three distinct scenarios. (i) In Section 3 we address the stationary case, i.e., a solution which does not decay in time. We focus on the resonance case, when the stationary solution resembles the non-accelerating bound state, and features a minimum degree of delocalization. (ii) In Section 4, we consider the particle originally trapped by the attractive delta-functional potential and dragged by it with the acceleration (the *pulling regime*). In the accompanying reference frame, this means a gradual tunnel decay of the original bound state. (iii) Section 5 deals with the case of a repulsive delta-functional potential, which is used to "push" the particle to the right, assuming that it was initially located on the right of the delta-function. In cases (ii) and (iii), which are illustrated below by Fig. 2, we aim to find the life time of the trapped state, and the largest velocity to which the particle can be accelerated, as long as it remains trapped (in terms of the guided-wave optics, imparting the "velocity" means deflecting the beam in a direction different from the original one, which is also an issue of an obvious interest to applications). Physical estimates for the largest velocity and life time are given for both the steered atoms and optical beams; in the latter case, the "life time" actually means the largest propagation distance. The paper is concluded by Section 6.

**2. The model**

**2.1. The laboratory reference frame**

In terms of quantum mechanics, our model is based on the attractive or repulsive δ-functional potential, which is quiescent at $t < 0$, and starts to run at a constant acceleration, $a$, at $t = 0$. In fact, any potential well supporting a single bound state, whose characteristic width is much larger than the well's width, may be approximated by the δ-function (see, e.g., Refs. [11-13]). The respective scaled Schrödinger equation, with $\hbar = m = 1$, is

$$i\frac{\partial \psi}{\partial t} = -\frac{1}{2}\frac{\partial^2 \psi}{\partial x^2} - \varepsilon \delta\left(x - \frac{a}{2}t^2 \theta(t)\right)\psi , \qquad (1)$$

where $\theta(t) = 0$ at $t < 0$ and $1$ at $t > 0$, and $\varepsilon > 0 / \varepsilon < 0$ is the strength of the attractive/repulsive potential.

In optics, Eq. (1) may be realized as the paraxial propagation equation for the amplitude of the electromagnetic wave in the spatial domain, with $t$ playing the role of the propagation distance (usually denoted $z$), while $x$ is the transverse coordinate. In that case, the delta-functional potential accounts for a narrow guiding stripe, which is an attractive channel or repulsive barrier, in the cases of $\varepsilon > 0$ and $\varepsilon < 0$, respectively. The acceleration is then realized as a bend of the guiding stripe in the $(x,z)$ plane.

To cast the original quantum-mechanical Schrödinger equation into the form of Eq. (1), we assume that $x$ is measured in units of $x_0 = 1$ nm. Then, for the typical case of the mass of rubidium atoms, the time unit in Eq. (1) is $t_0 = 1.5$ ns. Further, if the δ-function in Eq. (1) approximates a potential well of width $\Delta$ and depth $U_0$, the same rescaling gives $\varepsilon \equiv (t_0 U_0 \Delta)/(\hbar x_0)$. A natural size of the width is $\Delta \sim 1$ Å. Then, the scaled strength of the potential $\varepsilon \sim 1$ implies that the wave function of the bound state, $\phi(x) = \phi_0 \exp(-\varepsilon |x|)$, which corresponds to the energy eigenvalue $E_{\text{bound}} = -\varepsilon^2/2$, is localized over length $|x| \sim \varepsilon^{-1} \sim 1$ [see also Eq. (16) below]. In physical units, this localization length is $\sim 1$ nm, which is indeed much larger than the width of the well, thus justifying the use of the δ-functional potential. In physical units, the respective depth of the potential well is estimated as $U_0 \sim 10^{-5}$ eV. It is also relevant to mention that, for the known empiric size of the rubidium atom, $\approx 2.3$ Å, the assumed localization length, $x_0 = 1$ nm, is tantamount to $\sim 4$ atomic sizes, which is reasonable for potential applications.

In the application to the guided optical beams, a natural unit of the transverse coordinate is $x_0 = 1$ μm. The typical wavelength is on the same order of magnitude, $\lambda \sim 1$ μm, hence Eq. (1), if derived from the underlying wave equation, implies the unit of the propagation distance to be $2\pi x_0^2 / \lambda \sim 10$ μm. Further, assuming that the δ-function represents a stripe of width $\sim 1$ μm with a typical value of the change of the refractive index in it, $\delta n \sim 10^{-3}$, we conclude that the corresponding value of the strength of the delta-functional potential in Eq. (1) is $\varepsilon \sim 0.1$.

**2.2. The moving reference frame**

The analysis is made more convenient by rewriting Eq. (1) in the reference frame moving with acceleration $a$, i.e., in terms of the following variables [14] (in optics, this means the use of a curvilinear coordinate system in the plane of the waveguide):

$$\psi(x,t) \equiv \varphi(\xi,\tau) \exp\left(iaxt - \frac{i}{3}a^2 t^3\right), \qquad (2)$$

$$\xi \equiv 2\varepsilon\left(x - \frac{a}{2}t^2\right), \quad \tau \equiv 2\varepsilon^2 t. \qquad (3)$$

The accordingly transformed equation (1) is, at $\tau > 0$,

$$i\frac{\partial \varphi}{\partial \tau} = -\frac{\partial^2 \varphi}{\partial \xi^2} - \delta(\xi)\varphi + F\xi\varphi, \qquad (4)$$

where $\varepsilon > 0$ is assumed (here, we consider the case of the attractive δ-function), and the strength of the effective linear field induced by the acceleration is

$$F \equiv 2a/(2\varepsilon)^3. \qquad (5)$$

Seeking for a stationary solution to Eq. (4) in the form of $\varphi(\xi,\tau) = \phi(\xi)\exp(-iE\tau)$, the corresponding stationary equation is

$$\frac{d^2\phi}{d\xi^2} + \left[E - F\xi + \delta(\xi)\right]\phi = 0 \tag{6}$$

Solutions to Eq. (6) may be expressed in terms of Airy functions, $\text{Ai}(\zeta)$ and $\text{Bi}(\zeta)$ [15], with

$$\zeta \equiv F^{1/3}(\xi - E/F). \tag{7}$$

**3. Stationary solutions**

The well-known asymptotic form of the Airy functions at $\zeta \to +\infty$ is [15]

$$\text{Ai}(\zeta) \sim \tfrac{1}{2}\pi^{-1/2}\zeta^{-1/4}\exp\left[-(2/3)\zeta^{3/2}\right], \tag{8}$$

$$\text{Bi}(\zeta) \sim \pi^{-1/2}\zeta^{-1/4}\exp\left[(2/3)\zeta^{3/2}\right].$$

To avoid the divergence of the wave function in this limit, solutions are looked for as

$$\phi(\zeta) = \begin{cases} \text{Ai}(\zeta)/\text{Ai}(\zeta_0), & \zeta > \zeta_0, \\ [\text{Bi}(\zeta) + c\text{Ai}(\zeta)]/[\text{Bi}(\zeta_0) + c\text{Ai}(\zeta_0)], & \zeta < \zeta_0, \end{cases} \tag{9}$$

where the role of the renormalized energy eigenvalue is played by

$$\zeta_0 \equiv -E/F^{2/3}. \tag{10}$$

Ansatz (9) automatically satisfies the continuity condition at $\zeta = \zeta_0$, i.e., $\xi = 0$. The jump of the derivative at this point, following from the integration of Eq. (4) in an infinitesimal vicinity of $\xi = 0$, gives rise to the following condition:

$$F^{1/3}\left[\frac{\text{Ai}'(\zeta_0)}{\text{Ai}(\zeta_0)} - \frac{\text{Bi}'(\zeta_0) + c\text{Ai}'(\zeta_0)}{\text{Bi}(\zeta_0) + c\text{Ai}(\zeta_0)}\right] = -1 \tag{11}$$

(the prime stands for the derivative). Taking account of the fact that the Wronskian of the Airy functions is $W[\text{Ai}(\zeta_0), \text{Bi}(\zeta_0)] = 1/\pi$, we obtain from here

$$c = \frac{1}{\text{Ai}(\zeta_0)}\left[\frac{F^{1/3}}{\pi\text{Ai}(\zeta_0)} - \text{Bi}(\zeta_0)\right], \tag{12}$$

which means that the wave function can be written as

$$\phi(\zeta) = \frac{\text{Ai}(\zeta)}{\text{Ai}(\zeta_0)} + \begin{cases} 0, & \zeta > \zeta_0, \\ \pi F^{-1/3}[\text{Ai}(\zeta_0)\text{Bi}(\zeta) - \text{Bi}(\zeta_0)\text{Ai}(\zeta)], & \zeta < \zeta_0. \end{cases} \tag{13}$$

Clearly, this wave function is unnormalizable, due to the asymptotic form of the Airy functions at $\zeta \to -\infty$ [15],

$$\{\mathrm{Ai}(\zeta), \mathrm{Bi}(\zeta)\} \approx \pi^{-1/2}(-\zeta)^{-1/4} \{\sin, \cos\}\left(\frac{2}{3}(-\zeta)^{3/2} + \frac{\pi}{4}\right), \tag{14}$$

which corresponds to the following expression for the slowly decaying "tail" of the average probability density ($I$) in the region of $\zeta < \zeta_0$:

$$I \equiv \langle \phi^2 \rangle = \frac{1}{\pi\sqrt{-\zeta}}\left\{\left[\frac{1}{\mathrm{Ai}(\zeta_0)} - \frac{\pi}{F^{1/3}}\mathrm{Bi}(\zeta_0)\right]^2 + \left[\frac{\pi}{F^{1/3}}\mathrm{Ai}(\zeta_0)\right]^2\right\}. \tag{15}$$

The divergence of the total norm may be minimized in case the amplitude of the tail attains a minimum, which may be expected in a vicinity of the resonant energy ($E = -1/4$, in the present notation). It corresponds to the above-mentioned bound state supported by the attractive δ-function in the absence of the linear potential:

$$\varphi(\xi, \tau = 0) = (1/\sqrt{2})\exp(-|\xi|/2). \tag{16}$$

The minimum of the tail's amplitude, that defines an optimal quasi-bound (*least delocalized*) state in the presence of the linear potential, corresponds to the vanishing of the first term in the curly brackets in Eq. (15). Using the asymptotic approximation (8) for the case of small acceleration $a$, and taking definition (10) into regard, one can find that the minimum is attained at the value of the rescaled energy shifted from the aforementioned $E = -1/4$, which is valid at $a = 0$, to

$$E \approx -(1/4)(1 + 20F^2) \tag{17}$$

(in the original notation, it follows from Eq. (3) that this expression corresponds to $\tilde{E} \equiv 2\varepsilon^2 E = -(\varepsilon^2/2)[1 + 5a^2/(4\varepsilon^6)]$). The respective wave function is

$$\phi(\zeta) = \begin{cases} \mathrm{Ai}(\zeta)/\mathrm{Ai}(\zeta_0), & \text{at } \zeta > \zeta_0, \\ \mathrm{Bi}(\zeta)/\mathrm{Bi}(\zeta_0), & \text{at } \zeta < \zeta_0, \end{cases} \tag{18}$$

with tail (15) reducing to

$$I = \frac{1}{\pi\sqrt{-\zeta}}\left[\frac{\pi}{F^{1/3}}\mathrm{Ai}(\zeta_0)\right]^2 \approx \frac{1}{2\sqrt{-\zeta}F^{1/3}}\exp\left(-\frac{1}{6F}\right). \tag{19}$$

Profiles of the resonance (quasi-bound) states, as given by Eq. (18) with energy (17), are displayed in Fig. 1 for different values of effective field $F$ (the rescaled acceleration). The asymmetry of these plots and those displayed in other figures below, is a direct consequence of the effective linear potential replacing the acceleration. As might be expected, the localization is nearly ideal for small acceleration $a$, quickly deteriorating with the increase of $a$. For this reason, the search for the quasi-bound states at larger accelerations, when the above asymptotic analysis is not valid, does not really make sense.

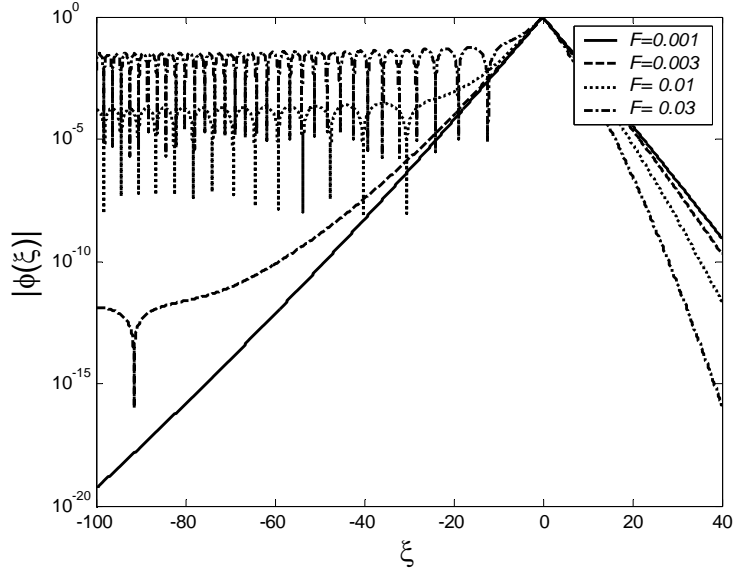

Fig.1: The absolute-value profiles of the wave function of the quasi-bound (least delocalized) stationary state are shown, on the logarithmic scale, versus coordinate ξ for different values of strength $F$ of the linear field [it represents the acceleration in the laboratory reference frame, see Eq. (5)].

In real media, as concerns the applications to the particle-transport systems and guided optical transmission alike, the situation may be affected by an environment in the form of a weak randomly scattering potential. This environment may tend to break the slowly decaying tail by gradually obliterating its coherence [16]. The limitation imposed by the fragmentation of the tail, caused by the random potential, on the above results depends on its strength and correlation length.

**4. The pulling scenario**

Proceeding to the consideration of dynamical settings, we start with the "pulling scenario", illustrated by the left panel in Fig. 2: the delta-functional potential well starts to move to the right, pulling the stationary state (16) originally trapped in it. In this case, the quasi-bound state presented in the previous section, which minimizes the delocalization of the wave function, cannot be formed from initial wave function (16), therefore the "pulling" scenario should be studied as a separate dynamical problem.

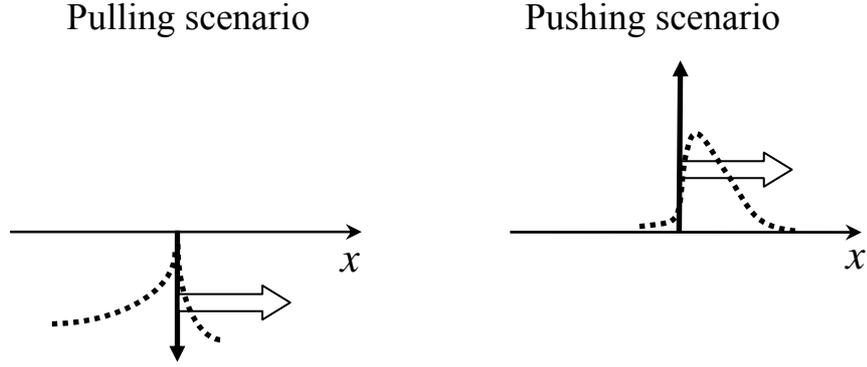

Fig. 2: The left and right panels illustrate the dynamical "pulling" and "pushing" scenarios. The vertical arrows designate the attractive and repulsive delta-functional potentials, and the right arrows indicate the direction of the acceleration. In each case, the dotted curves indicate qualitatively the profiles of the wavefunctions.

After the onset of the acceleration (at $\tau > 0$), the initial bound state will be transformed into a combination of evanescent and outgoing waves, i.e., the solution as a whole will be decaying in time. Using the same notation as above, i.e., Eqs. (2)-(5) and (10), the spatial part of the solution can be looked for as

$$\phi(\zeta) = \begin{cases} \text{Ai}(\zeta)/\text{Ai}(\zeta_0), & \text{at } \zeta > \zeta_0, \\ \left[\text{Bi}(\zeta) + i\text{Ai}(\zeta)\right]/\left[\text{Bi}(\zeta_0) + i\text{Ai}(\zeta_0)\right], & \text{at } \zeta < \zeta_0 \end{cases} \quad (20)$$

[cf. Eq. (9) for the stationary wave function], the combination $\left[\text{Bi}(\zeta) + i\text{Ai}(\zeta)\right]$ representing the outgoing wave, according to asymptotic expressions (14). However, in this case $E$ is a complex eigenvalue, whose imaginary part accounts for the temporal decay of the initially trapped state.

The jump in the derivative, following from the integration of Eq. (4) in an infinitesimal vicinity of the δ-function, gives rise, after a straightforward substitution, to the following condition, cf. Eq. (11):

$$F^{1/3} = \pi \text{Ai}(\zeta_0)\left[\text{Bi}(\zeta_0) + i\text{Ai}(\zeta_0)\right]. \quad (21)$$

A solution to this equation is a complex one, that can be defined as follows:

$$\zeta_0 = \zeta_0^{(R)} + i\zeta_0^{(I)} \equiv -(E_R + iE_I)/F^{2/3}, \quad (22)$$

see Eq. (10). Equation (21) can be solved numerically. In Fig. 3 we plot the real and imaginary parts of the solution versus the inverse driving field, $1/F$.

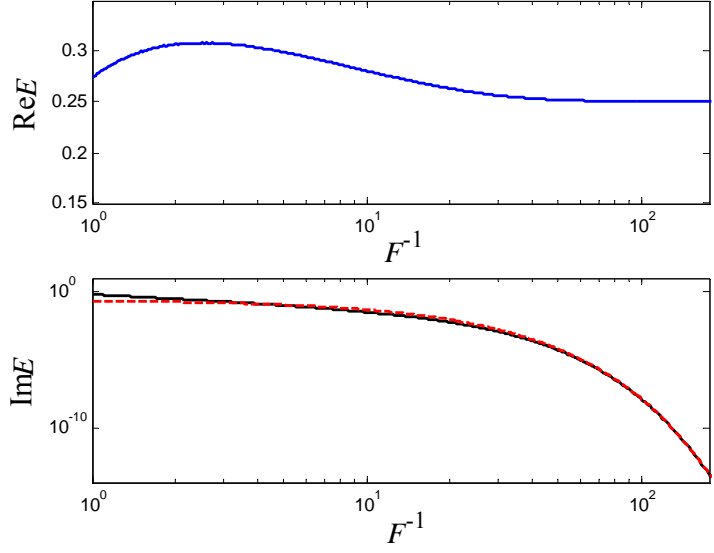

Fig. 3: (Color online) Real and imaginary parts of eigenvalue $E$, found from a numerical solution of Eq. (21), versus the inverse driving field, which is related to the acceleration in the laboratory reference frame as per Eq. (5). The dotted curve (red) in the lower panel stands for the weak-field approximation given by Eq. (26)

The imaginary part of the eigenvalue determines the probability of the decay of the accelerating bound state,

$$P \sim \exp(-2E_I \tau) \equiv \exp(-4E_I \varepsilon^2 t) \quad (23)$$

(recall $t$ and $\tau$ are related by Eq. (3)). Therefore, the lifetime of the bound particle in the trap moving with the constant acceleration may be estimated as

$$T = \left(4E_I \varepsilon^2\right)^{-1}, \quad (24)$$

which means that the maximum velocity the particle can be accelerated to is

$$v_{\max} \approx aT = a / \left(4E_I \varepsilon^2\right) \quad (25)$$

(as said above, at the second stage of the operation, the particle may be symmetrically decelerated back to the quiescent state, of the objective is to deliver the particle to a target location).

The initial ($t = 0$) profiles of the accelerating modes (20) with the numerically found eigenvalue (22) are plotted in Fig. 4 for various field strengths $F$ [i.e., for different accelerations $a$, see Eq. (5)]. A characteristic example of the corresponding spatiotemporal profile of the evolving probability density is displayed in Fig. 5.

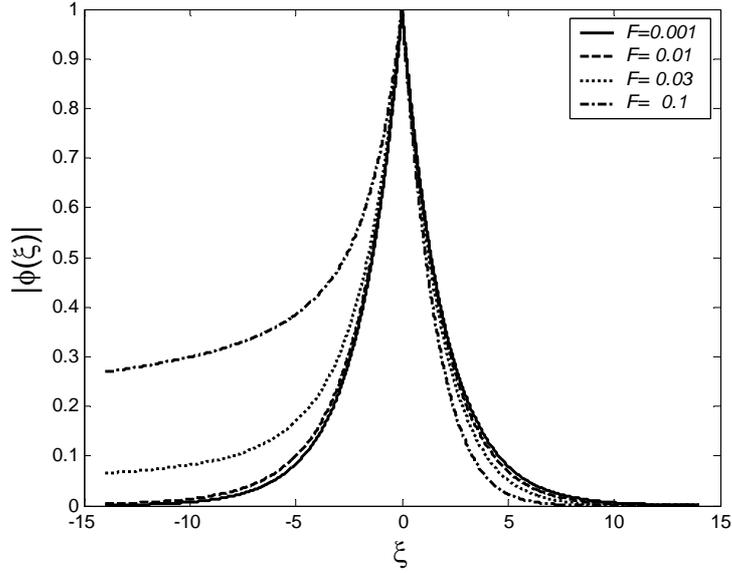

Fig. 4: The initial ($t = 0$) absolute value of the eigenmode (20) moving with the constant acceleration, as a function of coordinate $\xi$ for different values of the effective field $F$.

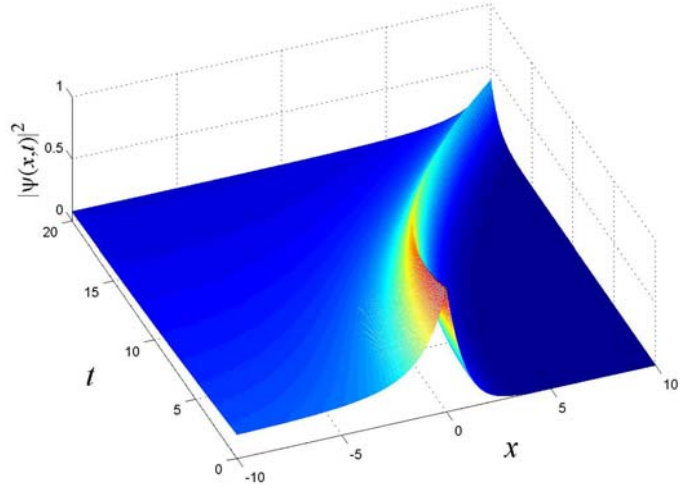

Fig. 5: (Color online) The spatiotemporal dynamics of the probability density for $F = 0.1$ and $\varepsilon = 1/2$, produced by Eqs. (20-22).

For $F \to 0$, $|\zeta_0| \gg 1$ is a large parameter as per Eq. (10), hence one can use the asymptotic approximation (8) to derive an analytical approximation for the complex root of Eq. (21):

$$E \approx -(1/4)\left[1 + i\exp(-1/6F)\right]. \tag{26}$$

In particular, the imaginary part of this expression corresponds to the life time of the trapped state, $\tau \sim 2\exp(1/6F)$, or, in terms of the original evolution variable $t$ and parameters $\varepsilon$ and $a$ [see Eqs. (3) and (5)],

$$T \sim \varepsilon^{-2} \exp(2\varepsilon^3/3a) . \qquad (27)$$

Accordingly, the maximum velocity to which the particle may be accelerated is then estimated as [cf. Eq. (25)]

$$v_{max} \sim a\varepsilon^{-2} \exp(2\varepsilon^3/3a), \qquad (28)$$

and the probability for the accelerating particle to remain in the bound state decays in time as

$$P(t) \sim \exp\left[-\varepsilon^2 t \exp(2\varepsilon^3/3a)\right]. \qquad (29)$$

In the opposite case of the large-acceleration regime, Eq. (21) may be approximated by $F^{1/3} = (i/4)\zeta_0^{-1/2}\exp(-4\zeta_0^{3/2}/3)$, hence the solution becomes

$$\zeta_0 \sim 4^{-2/3} e^{2\pi i/3} \ln(F), \qquad (30)$$

which correspond to the complex energy,

$$E_0 \approx -(F/4)^{2/3} \ln(F)(i\sqrt{3}-1)/2 . \qquad (31)$$

Accordingly, the probability of the decay is

$$P \sim \exp\left[-\sqrt{3}(F/4)^{2/3} \ln(F)\tau\right] \qquad (32)$$

or, in the original notation,

$$P \sim \exp\left[-0.54 a^{2/3} \ln(a\varepsilon^{-3}/4) t\right], \qquad (33)$$

and the corresponding maximum velocity is

$$v_{max} \sim Ta \approx -\frac{2\varepsilon(16F)^{1/3}}{\sqrt{3}\ln(F)} = -\frac{2(4a)^{1/3}}{\sqrt{3}\ln(a/4\varepsilon^3)} . \qquad (34)$$

Undoing the rescaling and getting back to physical units in the typical settings outlined above, the present results give the following estimates for the main results, in the characteristic case of $a \sim (2/3)\varepsilon^3$ [see Eqs. (27) and (28)]: for the steered particle (atom), the maximum achievable velocity is $v_{max} \sim 1$ m/s, the corresponding acceleration being $a \sim 10^8$ g. In terms of the steered optical beams, the deflection $\sim 1°$ (which is sufficient for all applications) may be attained over the transmission distance $\sim 1$mm. As concerns effects of the environment, such as the random potential discussed in the previous section, they are expected to be less significant for the dynamical regimes than for the static one, as it is less important what will eventually happen to parts of the wave function lost by the accelerating particle.

**5. The pushing scenario**

As shown in the right panel of Fig. 2, the *repulsive* δ-function [i.e., a potential barrier, with $\varepsilon < 0$ in Eq. (1)] may be used to "push" the wavepacket, which was initially placed on one side of the barrier. This model can emulate the "piston" created by an accelerating repelling laser beam shoving a Bose-Einstein condensate [17,18]. The respective Schrödinger equation at $\tau > 0$, rescaled as per Eqs. (3) and (5), but with $\varepsilon$ replaced by $|\varepsilon|$, is [cf. Eq. (4)]

$$i\frac{\partial \varphi}{\partial \tau} = -\frac{\partial^2 \varphi}{\partial \xi^2} + \delta(\xi)\varphi + F\xi\varphi . \qquad (35)$$

The substitution of $\varphi(\xi,\tau) = \phi(\xi)\exp(-iE\tau)$ yields the stationary version of Eq. (35):

$$\frac{d^2\phi}{d\xi^2} + \left[E - F\xi - \delta(\xi)\right]\phi = 0. \qquad (36)$$

If the potential barrier induced by the repulsive δ-function were infinitely tall, the quantum particle obeying Eq. (36) would remain in the domain of $\xi > 0$, with the stationary eigenfunction

$$\phi(x) = N \operatorname{Ai}(\zeta), \qquad (37)$$

subject to the respective boundary condition

$$\phi(\zeta = \zeta_0) = 0, \qquad (38)$$

where $\zeta$ and $\zeta_0$ are the same as in Eqs. (7) and (10) (in this case, $\zeta = \zeta_0$ is the location of the infinitely high potential wall). The respective normalization constant is defined so that $N^2 \int_{\zeta_0}^{\infty} |\operatorname{Ai}(\zeta)|^2 dx = 1$. In this approximation, the eigenvalue of the energy is determined, as per Eq. (38), by the first zero of function $\operatorname{Ai}(\zeta)$, which is located at

$$\zeta \approx -2.33811 \equiv \overline{\zeta}_0, \qquad (39)$$

hence, from equation $\zeta_0 \equiv -EF^{-2/3} = -2.33811$, one obtains

$$E = 2.33811F^{2/3}. \tag{40}$$

It is also straightforward to find the respective normalization factor, $N = 1.4261F^{1/6}$. Note that the energy is positive in this case, as seen from Eq. (40). If slope $F$ of the linear potential in Eq. (36) is small enough (which again corresponds to small acceleration $a$), the δ-function may indeed be regarded as a very opaque barrier, suggesting to seek for a solution with the energy close to eigenvalue (40).

Taking into regard also the component of the wave function at $\zeta < \zeta_0$, the compound stationary solution to Eq. (36), is looked for as follows, similar to the case of the attractive δ-function:

$$\phi(\zeta) = \begin{cases} N\,\mathrm{Ai}(\zeta), & \zeta > \zeta_0, \\ N_0\left[\mathrm{Bi}(\zeta) + i\,\mathrm{Ai}(\zeta)\right], & \zeta < \zeta_0. \end{cases} \tag{41}$$

The continuity condition at point $\zeta = \zeta_0$ yields $N\,\mathrm{Ai}(\zeta_0) = \left[\mathrm{Bi}(\zeta_0) + i\,\mathrm{Ai}(\zeta_0)\right]N_0$, and the jump condition for the derivative and this point amounts to relation $F^{1/3}\left[N\,\mathrm{Ai}'(\zeta_0) - N_0\left(\mathrm{Bi}'(\zeta_0) + i\,\mathrm{Ai}'(\zeta_0)\right)\right] = N\,\mathrm{Ai}(\zeta_0)$, cf. Eq. (11). Making use of these two equations, along with the aforementioned expression for the Wronskian of the Airy functions, we arrive at the same result as in Eq. (21), but with the opposite sign:

$$F^{1/3} = -\pi\,\mathrm{Ai}(\zeta_0)\left[\mathrm{Bi}(\zeta_0) + i\,\mathrm{Ai}(\zeta_0)\right], \tag{42}$$

and, further,

$$N_0 = N\frac{\mathrm{Ai}(\zeta_0)}{\mathrm{Bi}(\zeta_0) + i\,\mathrm{Ai}(\zeta_0)} = -\frac{F^{1/3}}{\pi}\frac{N}{\left[\mathrm{Bi}(\zeta_0) + i\,\mathrm{Ai}(\zeta_0)\right]^2} \approx \\ -\frac{F^{1/3}}{\pi}\frac{N}{\left[\mathrm{Bi}(\zeta_0)\right]^2} \approx -1.544F^{1/3}N \approx -2.2F^{1/2}. \tag{43}$$

Now, when root $\zeta_0$ of Eq. (42) is close to $\bar{\zeta}_0$ [see Eq. (39)], one can expand it as follows:

$$\zeta_0 = \bar{\zeta}_0 + F^{1/3} + i\gamma F^{2/3}, \tag{44}$$

where $\gamma \approx 1.534$ and, as said above, $F$ is treated as a small parameter. Then, with regard to Eq. (10), the imaginary part of the energy eigenvalue can be obtained from Eq. (44):

$\mathrm{Im}\{E\} = -\lambda F^{4/3}$, which corresponds to the decay probability

$$P \sim \exp\left(-2\gamma F^{4/3}\tau\right) = \exp\left(-4^{-1/3}\gamma\varepsilon^{-2}a^{4/3}t\right) \tag{45}$$

and the respective maximum velocity,

$$v_{\max} \sim aT = a/\left(4\varepsilon^2 \operatorname{Im}\{E\}\right) \approx \varepsilon^2\gamma^{-1}\left(a/4\right)^{-1/3}, \tag{46}$$

cf. Eqs. (27-29) in the case of the attractive δ-function. The general estimates of relevant physical parameters in the case of the pushing scenario turn out to be the same as in the case of pulling, which were given in the previous section.

On the other hand, in the extreme case of very small $a$, Eq. (46) demonstrates that the largest velocity achievable by means of the pushing scenario scales as $a^{-1/3}$, while in the framework of the pulling scenario $v_{\max}$ may be exponentially large, as per Eq. (28). This difference is explained by the fact that the barrier separating the narrow potential well from the free space in the latter case is much thicker than in the former situation, when the potential well is confined by the mild linear slope and the narrow delta-functional barrier. For the same reason, the decay of the initially trapped state is very fast in the case of the pushing scenario and large acceleration (large $F$), therefore we do not consider that case here.

A typical example of the spatiotemporal evolution of the wave function driven by the "pushing" repulsive δ-function is displayed in Fig. 6, in terms of the original coordinate and time, $x$ and $t$ (cf. Fig. 5 for the attractive setting). The initial (at $t = 0$) profiles of the propagating eigenstates are plotted in Fig. 7 for various slopes $F$ of the linear potential. These pictures are generated by Eqs. (41)-(43).

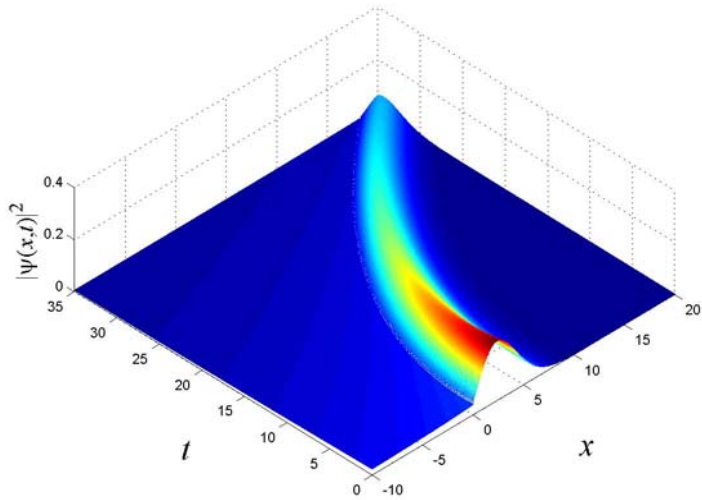

Fig. 6: (Color online) The spatiotemporal evolution of the probability density pushed by the repulsive δ-function., for $F = 0.06$ and $\varepsilon = 1/2$.

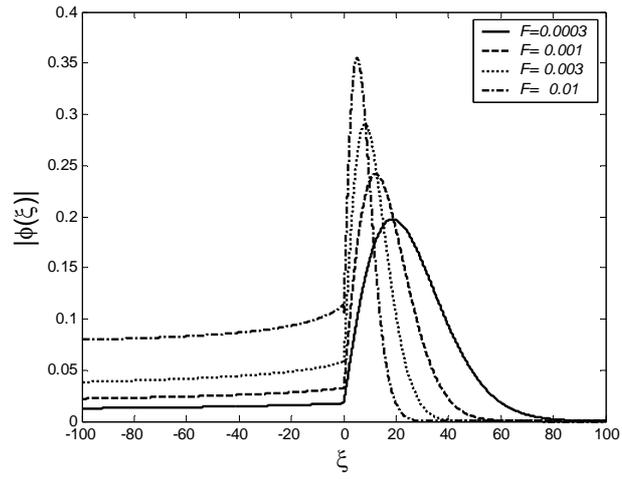

Fig. 7: The absolute value of the wave function in the initial eigenstates, given by Eqs. (41)-(43), pushed by the repulsive δ-function, for different values of the effective driving field $F$.

**6. Conclusion**

In this work, we have investigated the dynamics of the quantum particle in the presence of the potential of the attractive or repulsive δ-function moving with a constant acceleration. The consideration was performed in the reference frame moving along with the δ-function, which replaces the acceleration by the linear potential. The same setting may be realized in planar optical waveguides for beams steered by a narrow bending potential channel or barrier. Three main settings have been considered. (i) Stationary delocalized states of the resonance type in the case of the attractive δ-function, which minimize the degree of the delocalization, have been found. For the small acceleration, the deviation of the corresponding resonance energy from the energy of the bound state was found in the analytical form. (ii) The *pulling scenario*, for the quantum particle (or the optical beam) initially trapped in the bound state created by the attractive δ-function. In this case, the particle's life time in the bound state, and the largest velocity to which it may be accelerated before escaping, were also found in the analytical form, for a sufficiently small acceleration, as well as for large accelerations. Estimates were given for the corresponding physical parameters, in both cases of the steered atoms and guided optical beams. (iii) The *pushing scenario*, for the particle (or the beam) initially localized on the right of the repulsive δ-function, which moves to the right with constant acceleration. The largest velocity to which the particle can be accelerated before it escapes across the δ-function was found in this scenario too, for the case of a small acceleration.

An interesting extension of the present analysis may be to include nonlinearity (in the context of matter waves and optics alike). The self-attraction may help stabilize the bound state (a soliton) dragged by the accelerating potential trap. A challenging extension would be to consider dragging a two-dimensional bound state by a localized trap moving with acceleration in the two-dimensional plane.

**Acknowledgement**

The authors appreciate valuable discussions with Adolfo del Campo and Avi Marchewka.

FIGURES:

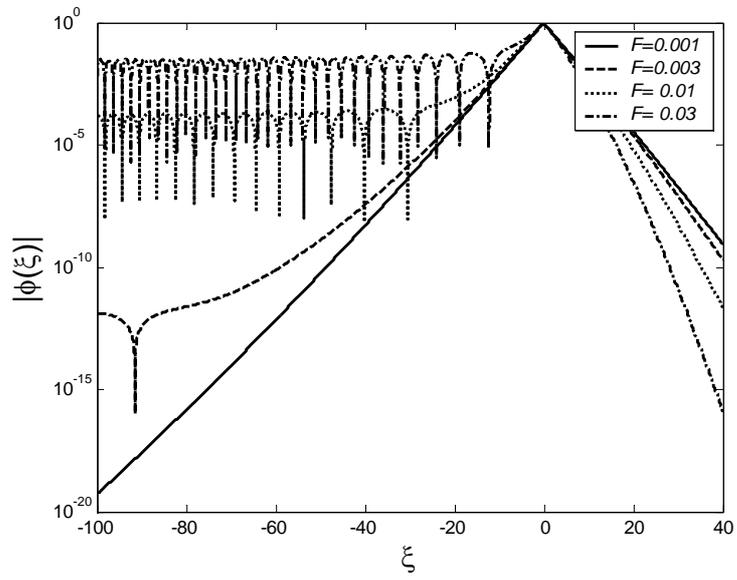

FIGURE 1

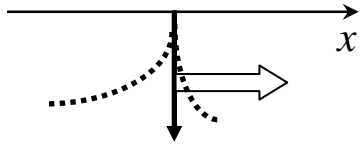 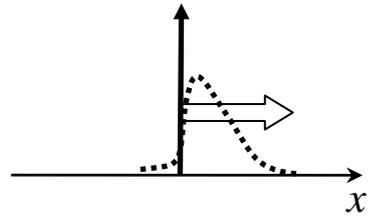

Pulling scenario   Pushing scenario

FIGURE 2

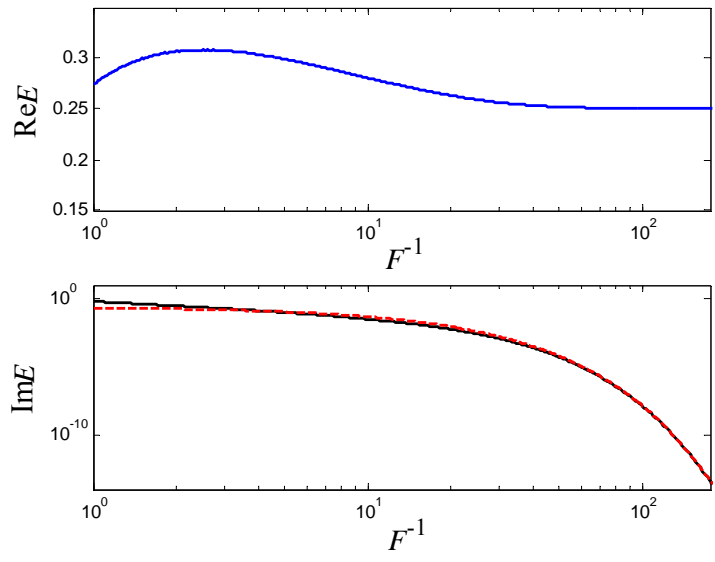

FIGURE 3

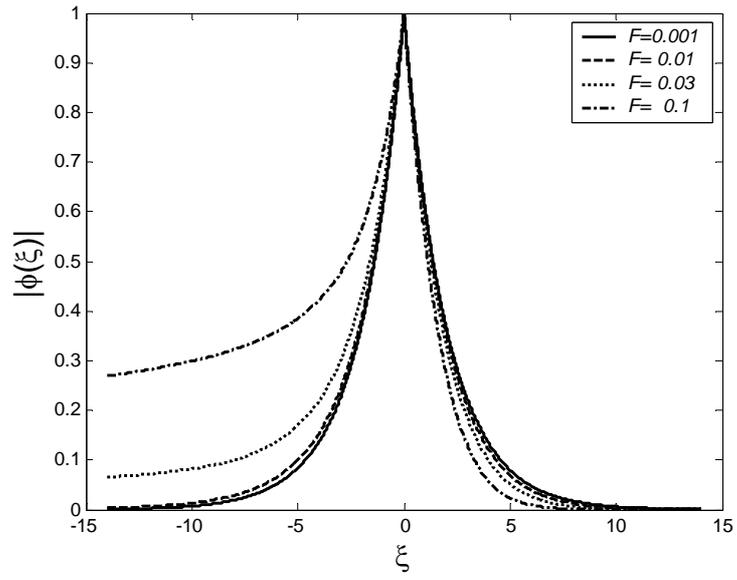

FIGURE 4

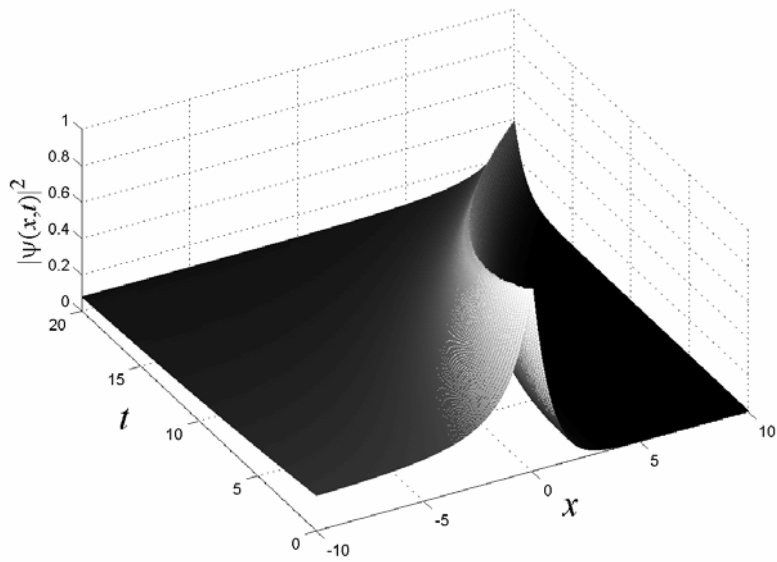

FIGURE 5

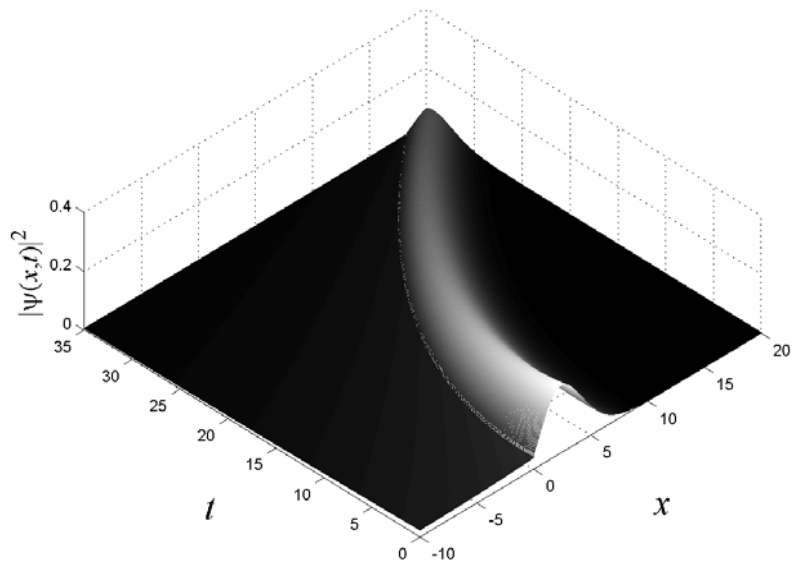

FIGURE 6

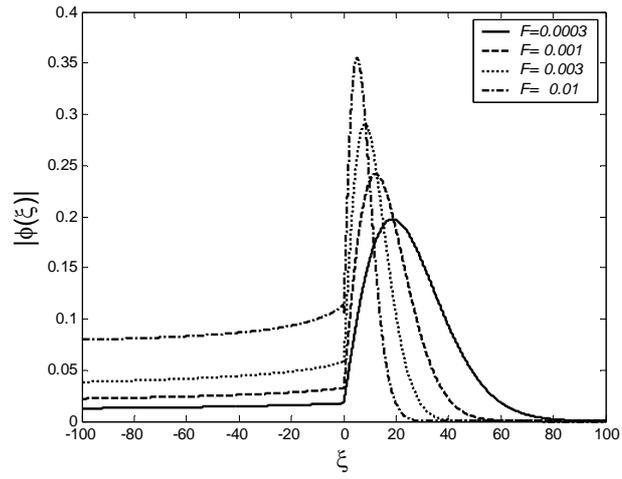

FIGURE 7

**Figure Captions**

Fig.1: The absolute-value profile of the wave function in the quasi-bound (least delocalized) stationary state is shown, on the logarithmic scale, as a function of coordinate $\xi$ for different values of strength $F$ of the linear field.

Fig. 2: The left and right panels illustrate the dynamical "pulling" and "pushing" scenarios. The vertical arrows designate the attractive and repulsive delta-functional potentials, and the right arrows indicate the direction of the acceleration. In each case, the dotted curves indicate qualitatively the profiles of the wavefunctions.

Fig. 3. (color online) Real and imaginary parts of eigenvalue $E$, found from a numerical solution of Eq. (21), vs. the inverse driving field $F$. The dotted curve in the lower panel stands for the weak- field approximation see Eq. (26)

Fig.4: The initial ($t=0$) absolute value of the wave function in the eigenmodes moving with the constant acceleration, as a function of coordinate $\xi$ for different values of the effective field $F$.

Fig.5: (color online) The spatiotemporal dynamics of the probability density for $F=0.1$ and $\varepsilon=1/2$.

Fig.6: (color online) The spatiotemporal evolution of the probability density pushed by the repulsive δ-function., for $F=0.06$ and $\varepsilon=1/2$.

Fig.7: The absolute value of the wave function of the eigenstates pushed by the repulsive δ-function, for different values of the effective driving field $F$.